\documentclass[a4paper, 12pt]{revtex4}
\usepackage[dvips]{graphicx}
\usepackage[]{indentfirst}
\usepackage[]{amsfonts}
\usepackage[]{amssymb}
\usepackage[]{bm}

\begin{document}

\title{Autonomous Learning by Dynamical Systems with Inertial or Delayed  Feedbacks}

\author{Pablo Kaluza}
	\affiliation{National Scientific and Technical Research Council \& Faculty of Exact and Natural Sciences, National University of Cuyo, Padre Contreras 1300, 5500 Mendoza, Argentina.} 

\author{Alexander S. Mikhailov} 
	\affiliation{Abteilung Physikalische Chemie, Fritz-Haber-Institut der Max-Planck-Gesellschaft, Faradayweg 4-6, 14195 Berlin, Germany.}
	
\date{\today}

\begin{abstract}
Dynamical systems can autonomously adapt their organization so that the required target dynamics is reproduced. In the previous Rapid Communication [Phys. Rev. E \textbf{90},030901(R) (2014)], it was shown how such systems can be designed using delayed feedbacks. Here, the proposed method is further analyzed and improved. Its extension to adaptable systems, where delays are absent and inertial feedbacks are instead employed, is suggested.  Numerical tests for three different models, including  networks of phase and amplitude oscillators, are performed.  
\end{abstract}


\maketitle

\section{Introduction}
\label{sec_introduction} 

In contrast to modern computers, information processing in animals and humans does not involve digital encoding and numerical computation. Instead, the analog mode is employed - the animal brain is essentially a dynamical system that can emulate various other dynamical systems in its environment and thus predict their behavior \cite{mikhailov_synergeticsI}. The question is whether one can construct dynamical systems that would be much more simple then neural networks, but nonetheless able to perform the tasks of emulation and prediction. 

The current study is a continuation of previous work on engineering of dynamical systems where evolutionary methods have been employed for systems design. In 1988, it was proposed that, by simulated annealing, oscillatory dynamical systems with prescribed frequency spectra can be built \cite{mikhailov_1988}; subsequently this was confirmed in numerical simulations \cite{Ipsen_mikhailov}. For flow distribution networks that represent an idealization of the signal transduction system of a biological cell, it was shown that, by using stochastic Metropolis algorithms, networks with prescribed static response patterns can be designed and made furthermore robust against various structural perturbations \cite{kaluza_pre2007} or noise \cite{kaluza_epl_2007} (see also review \cite{kaluza_chaos_2008}).
These results could be further extended to flow network models with dynamic responses \cite{kaluza_epjb_2012}. For coupled phase oscillators, it was shown that, by running evolutionary optimization, networks that are better entrained by pacemakers could be designed \cite{Kori_2006,Kaluza_book_2006}. 

Model genetic oscillatory networks with prescribed oscillation periods could be designed by evolutionary optimization \cite{Kobayashi_2007} and made robust against knock-outs of genes, removal of regulatory interactions, and introduction of noise \cite{Kobayashi_2011}. Moreover, genetic regulatory networks with predefined adaptive dynamical responses  \cite{Inoue_2013,Kaluza_Inoue_2016} or giving rise to definite stationary expression patterns  \cite{Fujimoto_2008} could be constructed in a similar way. Stochastic Monte Carlo optimization methods with replica exchange were employed to design phase oscillator networks with improved synchronization tolerance against heterogeneities \cite{Yanagita_2010}  or noise \cite{Yanagita_2012}. 

Application of evolutionary optimization methods in engineering of dynamical systems can be viewed as representing supervised learning: an external agent (actually, a digital computer)  generates structural perturbations (i.e., the mutations), monitors their effects on system's performance and decides whether a particular mutation should be accepted or not. In contrast to this, evolution leading to the designed dynamics should be performed entirely within the evolving system itself in the case of autonomous (or non-supervised) learning. Through an adaptation process, the system learns to adapt its internal structure in such a way that the desired performance is reached.

In our earlier Rapid Communication \cite{kaluza_pre2014}, a general scheme of autonomous learning by dynamical systems with delayed feedbacks has been proposed and illustrated using networks of coupled phase oscillators. Now, we provide a detailed analysis of the previously proposed method, consider its possible improvements and extensions and give further application examples.

\section{Methods}
\label{sec_methods}

We consider a system with variables $\boldsymbol{x} =(x_{1} ,x_{2} ,\ldots  ,x_{N})$ whose dynamics is governed by equations 

\begin{equation}
	\frac{d \boldsymbol{x}}{d t} =\boldsymbol{f} (\boldsymbol{x} ,\boldsymbol{w})
	\label{equ_intro_dyn_variables}
\end{equation} 

\noindent which depend on parameters $\boldsymbol{w} =(w_{1} ,w_{2} ,\ldots  ,w_{K})$. Suppose that this system is responsible in a selected application for execution of a function which needs to be optimized through the evolution of parameters $\boldsymbol{w}$. This function can be quantified by a performance vector $\boldsymbol{R} =(R_{1} ,R_{2} ,\ldots  ,R_{M})$ and, as we assume, this vector can be expressed as a function (or, generally, a functional) of system's variables, i.e. as $\boldsymbol{R} =\boldsymbol{F} (\boldsymbol{x})$. The ideal target performance of the system (\ref{equ_intro_dyn_variables}) is known and corresponds to a certain vector $\boldsymbol{R}^{\mathbf{0}}$. Thus, the deviation $\epsilon $ of the actual performance from the target performance of the system, i.e. the optimization error, can be defined as $\epsilon  =\vert \boldsymbol{R} -\boldsymbol{R}^{\mathbf{0}}\vert $ or, explicitly, as 

\begin{equation}
	\epsilon  =\vert \boldsymbol{F} (\boldsymbol{x}) -\boldsymbol{R}^{\mathbf{0}}\vert. 
	\label{equ_intro_error}
\end{equation} 

\noindent The error vanishes, $\epsilon  =0$, when the system reaches the target. 

Our aim is to introduce slow intrinsic dynamics of the parameters $\boldsymbol{w}$ which would lead to the minimization of error $\epsilon $. To do this, we assume that parameters $\boldsymbol{w}$ are themselves dynamical variables and their temporal evolution is governed by stochastic differential delay equations \cite{kaluza_pre2014} 

\begin{eqnarray}
	\frac{d w_{i} (t)}{d t} &  = &  -\frac{1}{\tau  (\epsilon  (t))} \left (\epsilon  (t) -\epsilon  (t - \Delta )\right ) \left (w_{i} (t) -w_{i} (t - \Delta )\right ) \nonumber  \\
	&  &  +S (\epsilon  (t)) \xi _{i} (t) .
	\label{equ_intro_dyn_parametros_delay}
 \end{eqnarray} 

\noindent where $\xi _{i} (t)$ are independent random white noises with $ \langle \xi  (t) \rangle  =0$ and $ \langle \xi _{\alpha } (t) \xi _{\beta } (t ' ) \rangle  =2 \delta _{\alpha  \beta } \delta  (t -t ' )$.  

Hence, the system consists of the guided subsystem (\ref{equ_intro_dyn_variables}) which is controlled by the slow steering subsystem (\ref{equ_intro_dyn_parametros_delay}). The
steering subsystem is persistently sensing the current and the previous performances of the guided subsystem and its deviations $\epsilon $ from the ideal target performance.

The evolution equation (\ref{equ_intro_dyn_parametros_delay}) has a simple interpretation: If, as a result of the parameter change $\delta  \boldsymbol{w} =\boldsymbol{w} (t) -\boldsymbol{w}(t - \Delta )$, the performance of the guided subsystem has improved, i.e. $\delta  \epsilon  =\epsilon  (t) -\epsilon  (t - \Delta ) <0$, the steering subsystem tends to change the parameters further in the same direction, increasing parameters $w_{i}$ if $\delta  w_{i} >0$ and decreasing them if  $\delta  w_{i} <0$. On the other hand, if the performance has become worse ($\delta  \epsilon  >0$), the steering subsystem moves the parameters in the opposite direction. 

After a parameter change, the guided subsystem would need some transient time to approach a new dynamic state. Only after that, the new performance of this subsystem can be evaluated. The delay $\Delta $ in equation (\ref{equ_intro_dyn_parametros_delay}) has to be chosen in such a way that it is larger then the characteristic transient time. On the other hand, this delay should be still smaller than the time within which the parameters of the system are varied. Suppose that the characteristic time scale of the guided subsystem (\ref{equ_intro_dyn_variables}) is equal to unity. Then, the condition $\Delta  \gg 1$ should be satisfied.

The function of the drift term in equation (\ref{equ_intro_dyn_parametros_delay}) is to correct the parameters in a deterministic way, based on previous performances and parameter values. But, if the evolution is deterministic, it may terminate in an intermediate optimum, before the target performance has been reached.  Therefore, a noise term is additionally included into this equation. The noise leads to stochastic exploration of the parameter space, i.e. to the diffusion process. As the target performance is gradually approached, the effects of noise should be however reduced. Therefore, we choose $S (\epsilon ) =S_{0} \epsilon  ,$ so that the noise strength is proportional to the deviation from the target performance. 

In the previous publication \cite{kaluza_pre2014}, we assumed that the parameter $\tau $ remains constant during the evolution, i.e. we have taken $\tau (\epsilon ) =\tau _{0} .$ Such choice is not however optimal because the error variation $\delta  \epsilon  =\epsilon  (t) -\epsilon  (t - \Delta )$ within the delay time can be much smaller than the error itself and,  as a result, diffusion can dominate over the drift. In the present study, we choose $\tau (\epsilon ) =\tau _{1}(\epsilon  +\mu )$ where $\mu  \ll 1$.  Hence, the drift becomes accelerated when the error $\epsilon $ is decreased. The small parameter $\mu $ is introduced to prevent the divergence when the target performance is reached and $\epsilon  =0$. 

In the considered learning scheme, there should be a separation of time scales between the guided and the steering subsystems. According to equation (\ref{equ_intro_dyn_parametros_delay}), the characteristic time for the evolution of the parameters $\boldsymbol{w}$ is $\tau (\epsilon )/\delta  \epsilon  =\tau _{1}(\epsilon  +\mu )/\delta  \epsilon $. Because, during the evolution, we typically have $\delta \epsilon  \ll \epsilon $, this implies that $\tau (\epsilon )/\delta  \epsilon  \gg \tau _{1}$. When the characteristic time scale for the dynamics of the guided subsystem is chosen as unity, the separation of time scale is, on the average, ensured if we choose the parameter $\tau _{1}$
to be of order one. 

As noted already in the Rapid Communication \cite{kaluza_pre2014}, the scheme employing fixed time delays may be sometimes difficult to implement at the hardware level, because, when it is employed,  the system needs to store in its memory the previous performances (i.e., the errors) and the previous parameter values. In such situations, learning schemes with inertial feedbacks can be used instead. 

The inertial feedback learning scheme is formulated by further extending the system, so that it includes an additional set of parameters $\boldsymbol{w}{\acute{}}$ and an auxilliary error variable $\epsilon {\acute{}}$. The evolution equations for parameters $\boldsymbol{w}$ are

\begin{equation}
\frac{d w_{i}}{d t} = -\frac{1}{\tau  (\epsilon )} \left (\epsilon  -\epsilon' \right ) \left (w_{i} -w'_{i}\right ) +S (\epsilon ) \xi _{i} (t),
\label{equ_intro_dyn_parametros_inercial}
\end{equation} 

\noindent and the dynamics of the additional variables is determined by equations  

\begin{equation} 
\Delta \frac{d w ' _{i}}{d t} =w_{i} -w'_i
\label{equ_intro_parametros_inercial}
\end{equation} 

and 

\begin{equation} 
\Delta \frac{d \epsilon  ' }{d t} =\epsilon  -\epsilon'. 
\label{equ_intro_error_inercial}
\end{equation} 

\noindent These equations involve a characteristic time for the additionally introduced variables that is taken to be the same as the previous delay time $\Delta $. The equation (\ref{equ_intro_dyn_variables}) for dynamical variables $x$ is not changed.  

To clarify the relationship between the inertial and delayed feedback versions, we note that equations (\ref{equ_intro_parametros_inercial}) and (\ref{equ_intro_error_inercial}) can be formally integrated, yielding

\begin{equation}
w'_{i} (t) =\int _{0}^{\infty }e^{ -\frac{s}{\Delta }} w_{i} (t -s) d s
\label{equ_intro_green_integral_w}
\end{equation} 

and 

\begin{equation}
\epsilon'(t) =\int _{0}^{\infty }e^{ -\frac{s}{\Delta }} \epsilon  (t -s) d s . 
\label{equ_intro_green_integral_e}
\end{equation} 

\noindent Therefore, for example, the auxilliary error $\epsilon'$ at time $t$ is obtained by summing contributions from different delayed moments $t -s$ taken with the exponentially decreasing weights $\exp ( -s/\Delta ) .$ Hence, in contrast to the previous scheme with a fixed delay $\Delta $, the inertial version effectively involves distributed time delays with the average equal to $\Delta $.

\section{Models}
To compare the operation and test the efficiency of different learning methods, three model problems will be used. Below we formulate these problems and specify the error functions and the parameter evolution equations in each case.

\textbf{A. Learning to maintain a given steady state.} Under strong environmental variations, the maintainance of a definite steady state may involve adaptation at the structural level, so that the parameters of a system become appropriately modified. As an illustration, we consider a simple problem of a system with gradient dynamics that learns to relax to a given stationary state.  

The system is characterized by dynamical variables $\boldsymbol{x} =(x_{1} ,x_{2} ,\ldots  ,x_{N})$ and parameters $\boldsymbol{w} =(w_{1} ,w_{2} ,\ldots  ,w_{K})$. The dynamical variables change with time according to equations 

\begin{equation}
\frac{d x_{i}}{d t} = -\frac{ \partial U (\bm x ,\bm w)}{ \partial x_{i}}. 
\label{equ_eje1_dynamics}
\end{equation} 

\noindent For simplicity, we assume that the potential function $U$ is given by the parabolic form 

\begin{equation}
U =\sum _{i =1}^{N}(x_{i} -w_{i})^{2} .
\end{equation}

Hence, the dynamics represents linear relaxation to a stationary state with $x_{i} =w_{i}$, i.e. we have 

\begin{equation}
\frac{d x_{i}}{d t} = -2 (x_{i} -w_{i}) .
\end{equation} 

We assume that variations of $x$ and the values of the parameters $w$ are restricted to the interval $( -R_{\max } ,R_{\max })$.

Suppose that the task of learning is to change the parameters so that a given target steady sate $\bm R^{0}$ is asymptotically reached. For this problem, the error function can be chosen as

\begin{equation}
\epsilon  (t) =\frac{1}{2NR_{\max }}\sum _{i =1}^{N}\left \vert x_{i} -R_{i}^{0}\right \vert  . 
\label{equ_eje1_error}
\end{equation} 

\noindent Thus defined, the error is always less than one. 

Then, the evolution of parameters $\bm w$ is described  by equations  (\ref{equ_intro_dyn_parametros_delay}) in the scheme with the delayed feedbacks and by equations (\ref{equ_intro_dyn_parametros_inercial}),  (\ref{equ_intro_parametros_inercial}) and (\ref{equ_intro_error_inercial}) under the inertial learning scheme.

Although the solution of this simple optimization problem can be easily constructed, it is still interesting to see how an autonomous learning system would deal with it.

\textbf{B. Learning to synchronize.} As our second example, we choose the problem that has already been taken in the previous Rapid Communication  \cite{kaluza_pre2014}. The system represents a network of coupled phase oscillators and it has to learn under what set of connection weights the state with a predefined synchronization level is achieved. 

The dynamical variables are the phases $\phi  =(\phi _{1} ,\phi _{2} , . . . ,\phi _{N})$ and the parameters are the connection weights  $w_{ij}$. The phase dynamics is described by the classical Kuramoto model \cite{Kuramoto}

\begin{equation}
\frac{d \phi _{i}}{d t} =\omega _{i} +\frac{1}{N} \sum _{j =1}^{N}w_{i j} \sin  (\phi _{j} -\phi _{i}) 
\label{equ_model_kuramotoI}
\end{equation} 

\noindent where $\omega _{i}$ is the natural frequency of an oscillator $i$.  We assume that the interactions are symmetric, $w_{i j} =w_{ji}$ and their weights can be positive or negative. 

The synchronization is quantified by the order parameter 

\begin{equation}
r (t) =\frac{1}{N} \left \vert \sum _{j =1}^{N}\exp  (i \phi _{j})\right \vert  . 
\label{equ_model_order_parameter}
\end{equation} 

\noindent Because it fluctuates with time, the degree of synchronization at time $t$ is better characterized by an average over a time interval $T$, 

\begin{equation}
R (t) =\frac{1}{T} \int _{t -T}^{t}r (t ' ) d t '  . 
\label{equ_model_mean_order_parameter}
\end{equation} 

\noindent Note that $R (t)$ can vary from zero to one and the value $R (t) =1$ corresponds to the state with complete phase synchronization, when the phases of all oscillators are identical. 

The task of learning in this case is to reach a structural state where synchronization at an arbitrary predefined level $P$ with $0 <P <1$ is achieved. This has to be done by only rearranging the weights (and possibly changing their signs), whereas the total interaction strength $\sum _{i ,j =1}^{N}\vert w_{i j} (t)\vert $ remains fixed.

The error $\epsilon $ can be defined as

\begin{equation}
\epsilon  =\vert P -R (t)\vert  . 
\label{equ_model_error}
\end{equation} 

When fixed delays are employed, the evolution equations for the connection weights are \cite{kaluza_pre2014}

\begin{eqnarray}
\frac{d w_{i j} (t)}{d t} &  = &  -\frac{1}{\tau  (\epsilon  (t))} \left (\epsilon  (t) -\epsilon  (t - \Delta )\right )(w_{i j} (t -T) \nonumber  \\
 &  &  -w_{i j} (t -T - \Delta )) +\lambda  \frac{w_{i j}(t)}{v (t)} \left (W -v (t)\right ) \nonumber  \\
 &  &  +S (\epsilon  (t)) \xi _{i j} (t) . 
 \label{equ_model_caso_delayed}
 \end{eqnarray} 

Here, a term is added to the right hand side of the equations to control the total strength of the interactions.  The average absolute weight $v (t)$ is defined as 

\begin{equation}
v (t) =\frac{1}{N (N -1)} \sum _{i ,j =1}^{N}\vert w_{i j} (t)\vert  
\label{equ_model_peso_medio_absoluto}
\end{equation} 

\noindent and this average weight should stay close to a given value $W$. The magnitude of the coefficient $\lambda $ determines how strictly this condition should hold. In contrast to the previous study, the dependences $\tau  (\epsilon ) =\tau _{1} (\epsilon  + \mu )$ and $S (\epsilon ) =S_{0} \epsilon $ will  now be used. 

When learning with inertial feedbacks is employed, evolution of connections weights is governed by equations

\begin{eqnarray}
\frac{d w_{i j}}{d t} &  = &  -\frac{1}{\tau  (\epsilon )} \left (\epsilon  -\epsilon  ' \right )(w_{i j} (t -T) -w ' _{i j}) \nonumber  \\
 &  &  +\lambda  \frac{w_{i j}}{v (t)} \left (W -v (t)\right ) +S (\epsilon ) \xi _{i j} (t) , 
 \label{equ_model_caso_inercial}
 \end{eqnarray} 

\begin{equation} 
\Delta \frac{d w ' _{i j}}{d t} =w_{i j} (t -T) -w ' _{i j}
\label{equ_model_pesos_delayed}
\end{equation} 

and the dynamics of the auxilliary error variable $\epsilon'$ is given by

\begin{equation} 
\Delta \frac{d \epsilon  ' }{d t} =\epsilon  -\epsilon  '  
\label{equ_model_error_delayed}
\end{equation} 

The same definition of $v (t)$ and the same dependences $\tau  (\epsilon )$ and $S (\epsilon )$ are then used. Note that the scheme still involves a short delay $T$ needed for the collection of data to determine the running time-averaged synchronization order parameter $R(t)$.

\textbf{C. Learning to avoid the amplitude death.} As the third example, a Kuramoto-Suzuki network of interacting amplitude oscillators is chosen. Previous investigations have shown \cite{Nakao2009} that such networks possess a rich dynamics. Particularly,  amplitude death for a fraction of oscillations in networks with random connections could be observed. The task of learning may be to modify the structure of a network so that the fraction of oscillators undergoing the ampltude death is minimized. 

In this model, the state of an oscillator $j$ is described by its complex amplitude $W_{j}$ and the oscillator dynamics is governed by equations

\begin{eqnarray}
	\frac{dW_j(t)}{dt} &  = & (1 +i c_{0}) W_{j} -(1 +i c_{2}) \vert W_{j}\vert ^{2} W_{j} \nonumber  \\
	&  &  +\kappa  (1 +i c_{1}) \sum _{k =1}^{N}L_{j k} W_{k} . 
	\label{equ_KT_dynamics}
 \end{eqnarray} 

\noindent where the coefficients $c_{0}$, $c_{1}$ and $c_{2}$ determine properties of individual oscillators and the parameter $\kappa $ specifies the strength of coupling between them. 

The coupling is characterized by the Laplacian matrix $\mathbb{L}$ with the elements 

\begin{equation}
	L_{jk} = w_{jk}T_{jk} -\delta _{j k} \sum _{l =1}^{N} w_{l j} T_{lj}. 
	\label{equ_KT_Laplaciam}
\end{equation} 

\noindent Here, $\mathbb{T}$ is the adjacency matrix of the network whose elements are $T_{i j} =1$ if there is a link from $j$ to $i$, and $T_{i j} =0$ otherwise. If a link exists, it is further characterized by its positive weight $w_{i j}$. We assume that the adjacency matrix represents a random Erd{\"o}s-R{\'e}nyi symmetric network with connectivity $p$. It will be fixed during the evolution and only the connection weights $w_{i j}$ will be changed.  

When the weights are all equal to unity, this system was previously investigated in Ref. \cite{Nakao2009}.  It was found that, under certain conditions, some of the oscillators in the system undergo the amplitude death, i.e their oscillation amplitude drops down to (almost) zero. Our aim is to distribute the connection weights over a given random network in such a way that the number of oscillators  with the amplitude death is reduced. 

We define the fraction $f$ of oscillators with the oscillator death as

\begin{equation}
	f (t) =\frac{1}{N} \sum _{j =1}^{N}\Theta  (\vert W_{j} (t)\vert ^{2} -h) . 
	\label{equ_KT_fraction}
\end{equation} 

\noindent where $h$ is a fixed threshold and $\Theta (x)$ is the step function, i.e. $\Theta (x) =1$ for $x >0$ and $\Theta (x) =0$ otherwise. 

This fraction fluctuates with time and, to define the error $\epsilon (t)$, averaging over a time interval $T$ is performed, 

\begin{equation}
	\epsilon (t) =\frac{1}{T} \int _{t -T}^{t}f (t ' ) d t '  . 
	\label{equ_KT_error}
\end{equation} 

To ensure that the connection weights remain positive during the evolution, the following procedure is applied:~We introduce additional variables $g_{i j}$, such that $w_{i j} =g_{i j}^{2}$. Thus, the weights stay positive whereas the variables $g$ can change their signs. With this convention,  the previous schemes for the autonomous learning can be applied.

We require that the average connection weight,

\begin{equation}
	v (t) =\frac{1}{\mathcal{L}} \sum _{i ,j =1}^{N} g_{ij}^{2}(t)T_{ij}
\end{equation}

\noindent is approximately conserved during the evolution and remains close to the unity. Here, $\mathcal{L}$ is the total number of connections in the considered fixed random network.

In the inertial learning scheme, the evolution of the variables $\bm g$ is governed by stochastic equations

\begin{eqnarray}
	\frac{d g_{i j} (t)}{d t} &  = &  -\frac{1}{\tau  (\epsilon )} \left (\epsilon  (t) -\epsilon  ' \right )(g_{i j} (t -T) -g ' _{i j}) \nonumber  \\
	&  &   +\lambda  g_{i j} (t) \left (1 -v (t)\right ) +S (\epsilon  (t)) \xi _{i j} (t) .
	\label{equ_KT_dyn_pesos_inertial}
 \end{eqnarray} 

\noindent Note that, similar to our second model, we have included into these equations a term that imposes the approximate conservation of the average connection weight $v (t)$.  The parameter $\lambda $ determines how strictly this condition is satisfied.

The evolution for the additional network variables $g'_{ij}$  and for the auxilliary error variable $\epsilon'$ are 

\begin{equation}
\Delta \frac{d g' _{ij}}{d t} =g_{ij} (t -T) -g'_{i j} 
\label{equ_KT_pesos_inertial}
\end{equation} 

and 

\begin{equation} 
\Delta \frac{d \epsilon  ' }{d t} =\epsilon  (t) -\epsilon  '  . 
\label{equ_KT_error_inertial}
\end{equation}

\section{Results}
\label{sec_numerical} 

In this section,  results of numerical simulations of autonomous learning for three problems, that were formulated above, will be presented and analyzed.

\textbf{A. Learning to maintain a given steady state.} This example will be used to discuss the effects of the dependence $\tau (\epsilon )$ and to compare the efficiency of the delay and inertial schemes. In the simulations, we take a system with $N =5$ variables and the target $R_{i}^{0} =10$ for $i =1 ,2 , . . . ,5$. The initial conditions for variables $x_{i}$ are randomly chosen within the interval $[ -R_{m a x} :R_{m a x}]$, with $R_{max}= 30$. In the delay scheme, the initial conditions for the parameters an their memory are $w_{i} = x_{i}$. In the inertial scheme, we have $w_{i} = w'_{i} =x_{i}$ and $\epsilon' =\epsilon$ at $t =0$. In both cases the initial errors are computed with the initial configuration of the system. The delay parameter is $\Delta  =10$ and $S_{0} =1$. The system is integrated numerically with a stochastic Heun algorithm with the time step of $0.001$.   

Typical simulation results are displayed in Fig. 1. Here, the dependence of error $\epsilon $ on time is shown for the delayed and inertial feedback schemes with $\tau (\epsilon ) =\tau _{0}$ (Fig. 1 a and b) or $\tau (\epsilon ) =\tau _{1}(\epsilon  +\mu )$ (Fig. 1 c and d). To facilitate the comparison, the same pseudo-random number series are used in all these simulations, so that the noise realizations are identical in all of them. Moreover, the same initial conditions are employed. We have $\tau _{0} =0.18 ,\tau _{1} =1 ,$ and $\mu  =0.0001$.

\begin{figure}[!ht] 
	\begin{center}
			\includegraphics[width=0.75\columnwidth,clip]{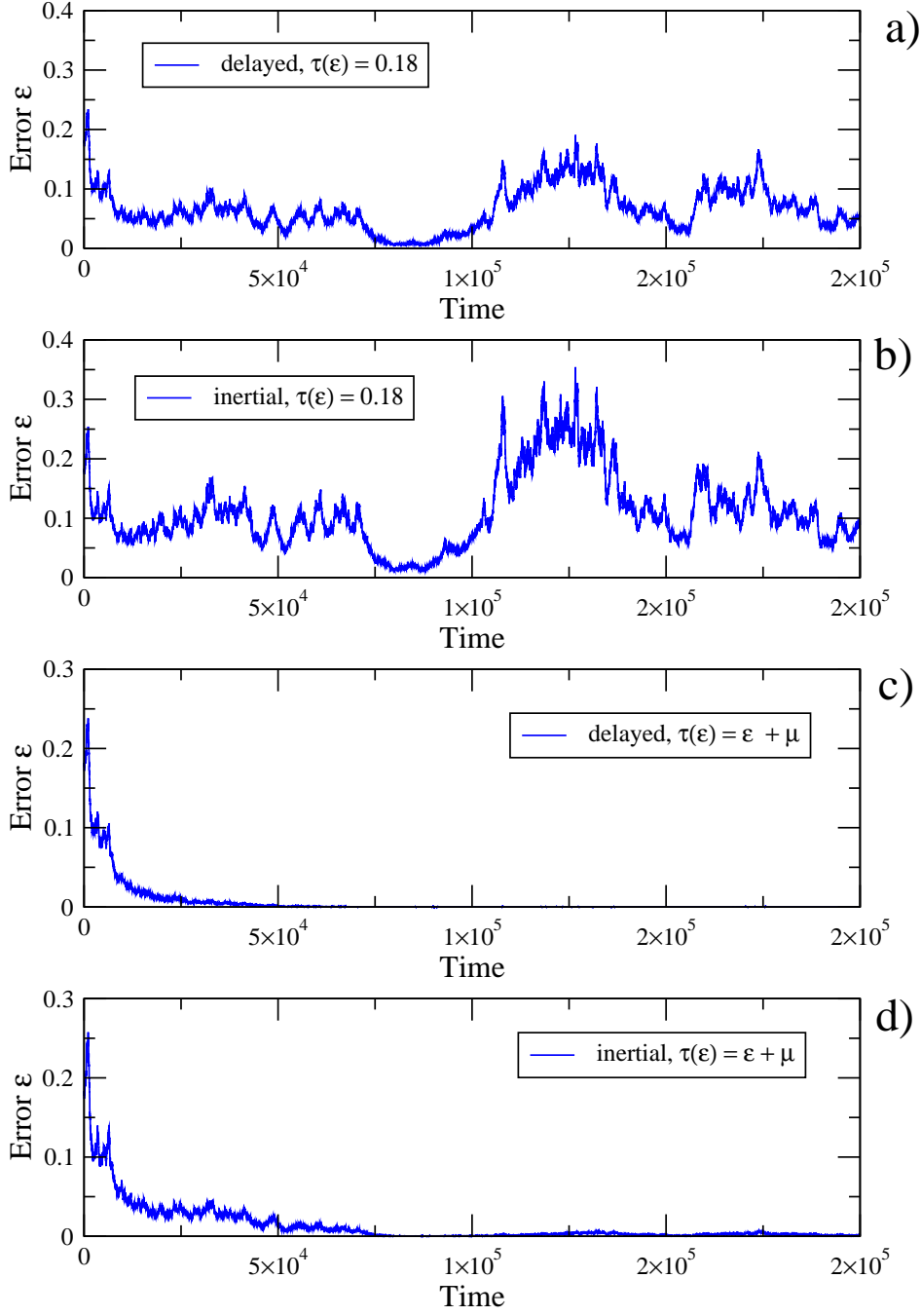}
			\caption{Comparison between the performances of different learning methods in model A:  (a) delayed scheme with $\tau(\epsilon) = 0.18$, (b) inertial scheme with $\tau(\epsilon) = 0.18$, (c) delayed scheme with $\tau(\epsilon) = \epsilon + \mu$, and, (d) inertial scheme with $\tau(\epsilon) = \epsilon + \mu$. The same initial conditions and the same realizations of the noise are used in all shown evolutions.}
			\label{fig_errores_4_casos} 
		\end{center}
	\end{figure} 

When the parameter $\tau $ remains constant during the evolution, in both schemes the system finds a relatively good solution at the time about $ 8 \times 10^{4}$. However, after that the errors grow again and the system moves away from the target performance. This behavior is repeated again and again, so that the target performance is never stabilized. Obviously, such bouncing is caused by the effects of noise that become dominant when the errors are small. In contrast to this, the errors gradually decrease with time when the dependence $\tau (\epsilon ) =\tau _{1}(\epsilon  +\mu )$ is employed and the target performance is asymptotically reached (Fig. 1c,d). The convergence to the target performance is somewhat better for the scheme with the delayed feedbacks.

\begin{figure}[!ht] 
		\begin{center}
			\includegraphics[width=0.75\columnwidth,clip]{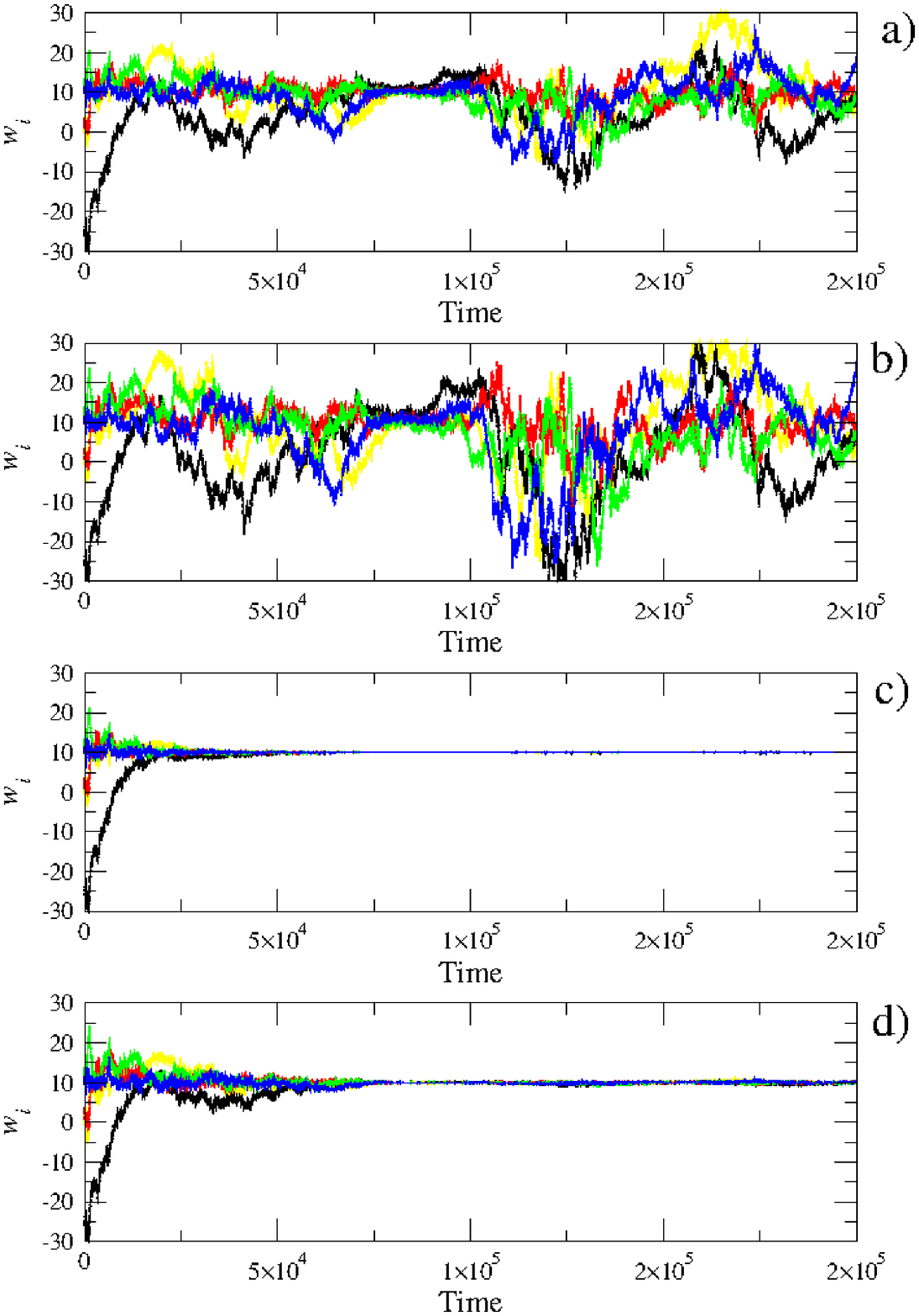}
			\caption{Evolution of parameters $(w_1, w_2,..., w_{10})$ under different learning schemes in model A. The same parameters and conditions as in Fig. \ref{fig_errores_4_casos}.}
			\label{fig_pesos_4_casos} 
		\end{center}
	\end{figure}

These differences are further illustrated by Fig. 2 where the dependence of the weights on the time for the same four evolutions is shown.  When $\tau  (\epsilon ) =\tau _{0}$, the weights approach the target values, but they cannot keep such values and the system soon escapes from the optimal state. On the other hand, when $\tau  (\epsilon ) =\tau _{1}(\epsilon  +\mu )$, the target values are found faster and the system can keep these values after they have been reached.

\textbf{B. Learning to synchronize.} In our Rapid Comminication \cite{kaluza_pre2014}, simulation results for this problem have been reported for the original delay scheme and $\tau (\epsilon ) =\tau _{0}$. Now we show how learning proceeds in this model if the dependence $\tau  (\epsilon ) =\tau _{1}(\epsilon  +\mu )$ is chosen instead, both in the delay and the intertial schemes. 

The system consists of $N =10$ coupled phase oscillators. Their natural frequencies $\omega _{i}$ are uniformly distributed between $ -0.3$ and $0.3$, so that $\omega _{i} =(i -1)/15 -0.3$ for $i =1 ,2 ,\ldots  ,10$. If uniform coupling with constant weights $w_{i j} =0.3$ is chosen, simulations show that the system with such parameters exhibits phase synchronization with $R  \approx 0.3$. 

The delay is  $\Delta  =100$ and the averaging interval is $T =50$. Moreover, we have $\tau _{1} =20$, $\mu  =0.05$, $S_{0} =0.01$, and $\lambda  =0.1$. The target order parameter is fixed at  $P =0.6$ and $W =0.3$. The equations are integrated using  the Euler algorithm with the constant time step $d t =0.01$. As initial conditions, we set all the weights and memory as $w_{ij} = w'_{ij} = W$; and the initial errors $\epsilon = \epsilon'$ is computed with the initial system configuration. Finally,  initial oscillator phases are randomly chosen between zero and $2\pi$.

\begin{figure}[!ht] 
		\begin{center}
			\includegraphics[width=0.75\columnwidth,clip]{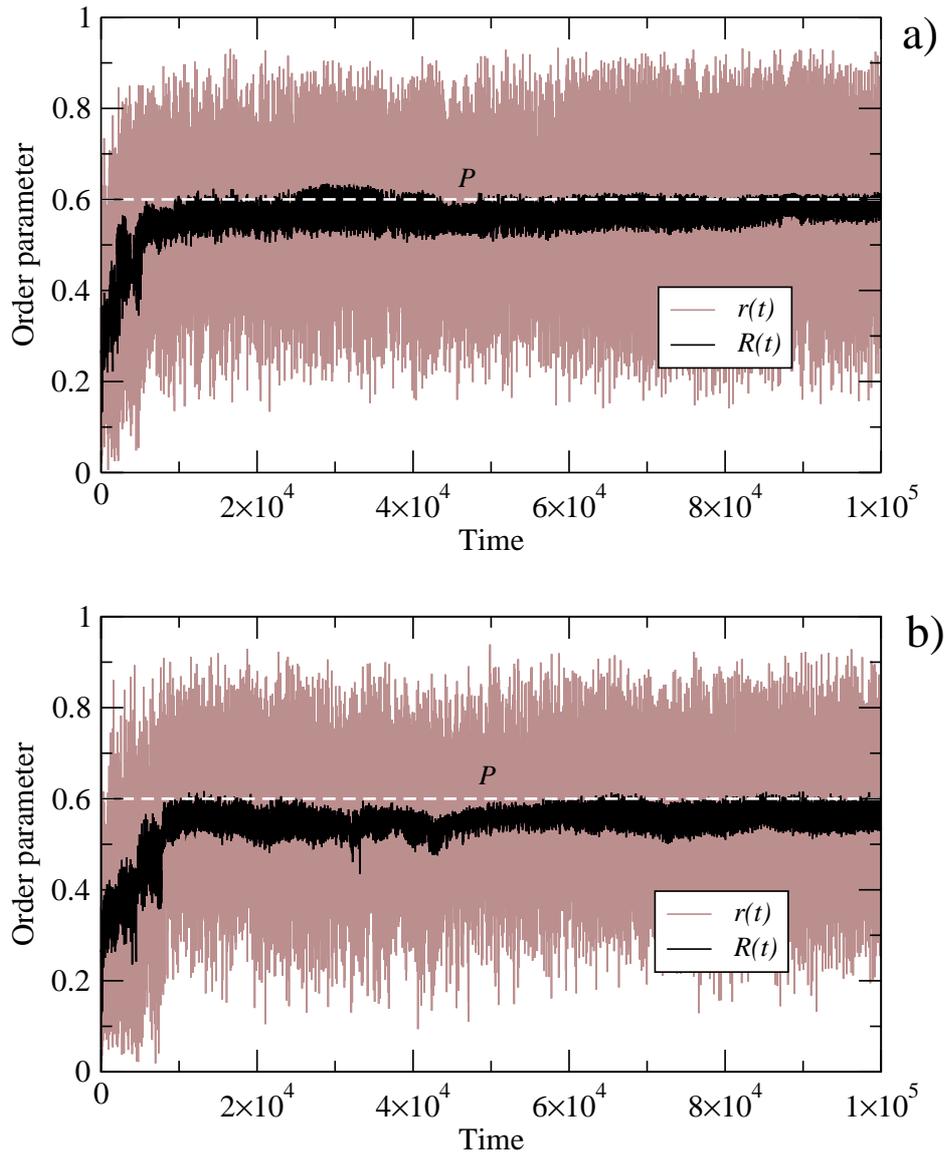}
			\caption{Evolution of the time-averaged synchronization order parameter $R(t)$ (black curve) in model B under (a) delayed and (b) inertial feedback schemes. The target value for synchronization is $P = 0.6$. Additionally, time dependences of the instantaneous order parameter $r(t)$ are shown for both schemes. } 
			\label{fig_kuramoto1_inercial} 
		\end{center}
	\end{figure}

As seen in Fig. 3 where the dependence of the synchronization order parameter $R$ on time is plotted, the system can learn to reproduce the required synchronization level of $R =0.6$ within the time about 10\textsuperscript {4} and approximately keeps this state afterwards under both learning schemes. The remaining fluctuations in $R(t)$ are due to intrinsic noise of the guided subsystem of phase oscillators. Such fluctuations appear much stronger if the instantaneous synchronization parameter $r(t)$ is displayed. The convergence to the target performance is again slightly faster for the scheme with delayed feedbacks. 

The final average values of the absolute weights are $v (t) =0.3$ at the end of the evolution for both schemes. Initially, the oscillators were globally coupled with the same weight $W =0.3$. Hence, the system learns to redistribute the weights so that the target functionality is reached.

\textbf{C. Learning to avoid the amplitude death.} We consider a fixed random network with $N =20$ oscillators and $37$ links that are generated with the connection probability $p =0.2$. The amplitude oscillators have parameters $c_{0} =0$, $c_{1} = -2$, $c_{2} =2$ and $\kappa  =0.25$. Our aim is to reduce the number of oscillators with the oscillation amplitude below $h =0.4$ by redistributing the connection weights over the given network. The parameters of the steering subsystem are $S_{0} =0.025$, $\lambda  =0.01$ and $\Delta  =100$. The inertial learning scheme is applied and we have  $\tau  (\epsilon ) =\tau _{1} (\epsilon  +\mu )$ with $\tau _{1} =10$ and $\mu  =0.01$. To evaluate the errors according to equation (\ref{equ_KT_error}), we set $T =50$. The system is integrated using the Euler method with $d t =0.01$. The initial conditions for the complex phases $W_j$ are random points over the unit circle on the complex plane. The initial weights and delayed weights satisfy $w'_{ij} = w_{ij} = 1$ for 
the elements with $T_{ij} = 1$. Finally, the errors $\epsilon' = \epsilon$ at $t=0$ are computed using the initial values.

\begin{figure}[!ht] 
	\begin{center}
		\includegraphics[width=0.75\columnwidth,clip]{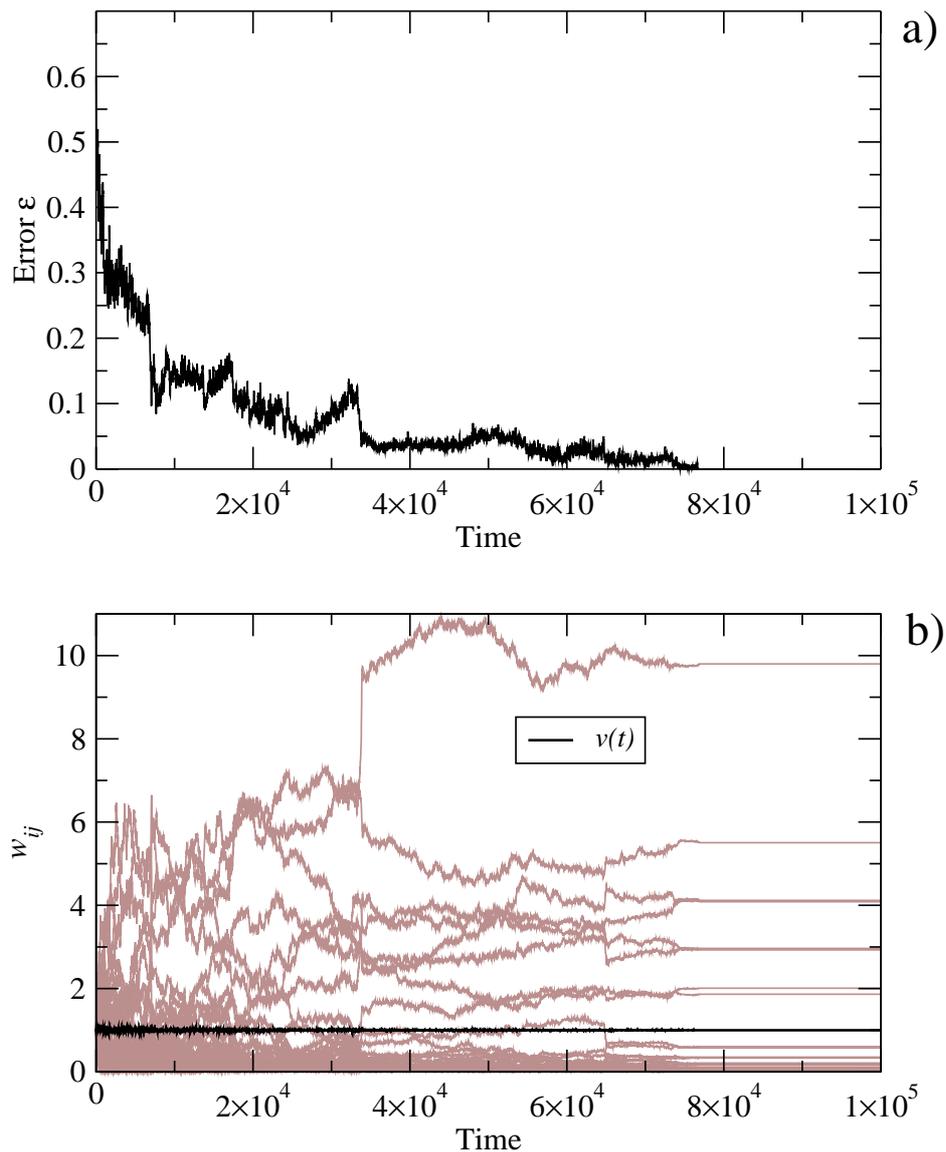}
		\caption{Evolutions of the error (a) and connection weights (b) in the random Kuramoto-Tsuzuki network for model C. The inertial learning scheme is applied. The black curve in (b) shows the dependence of the mean connection weight. } 
		\label{fig_KT_error} 
	\end{center}
\end{figure}

Figure \ref{fig_KT_error}a shows temporal evolution of the error $\epsilon $. While initially about $50 \%$ of the oscillators have the amplitudes inside the circle of radius $\sqrt{h}$ , at the end of the simulation this fraction is almost zero. The required structural state of the system is reached through the redistribution of connection weights. As seen in Fig. \ref{fig_KT_error}b, the weights $w_{ij}$ of individual connections vary largely during the evolution and the network becomes strongly heterogeneous in the final state. However, the mean connection weight  $v (t)$ remains close to one.

\begin{figure}[!ht] 
		\begin{center}
			\includegraphics[width=0.75\columnwidth,clip]{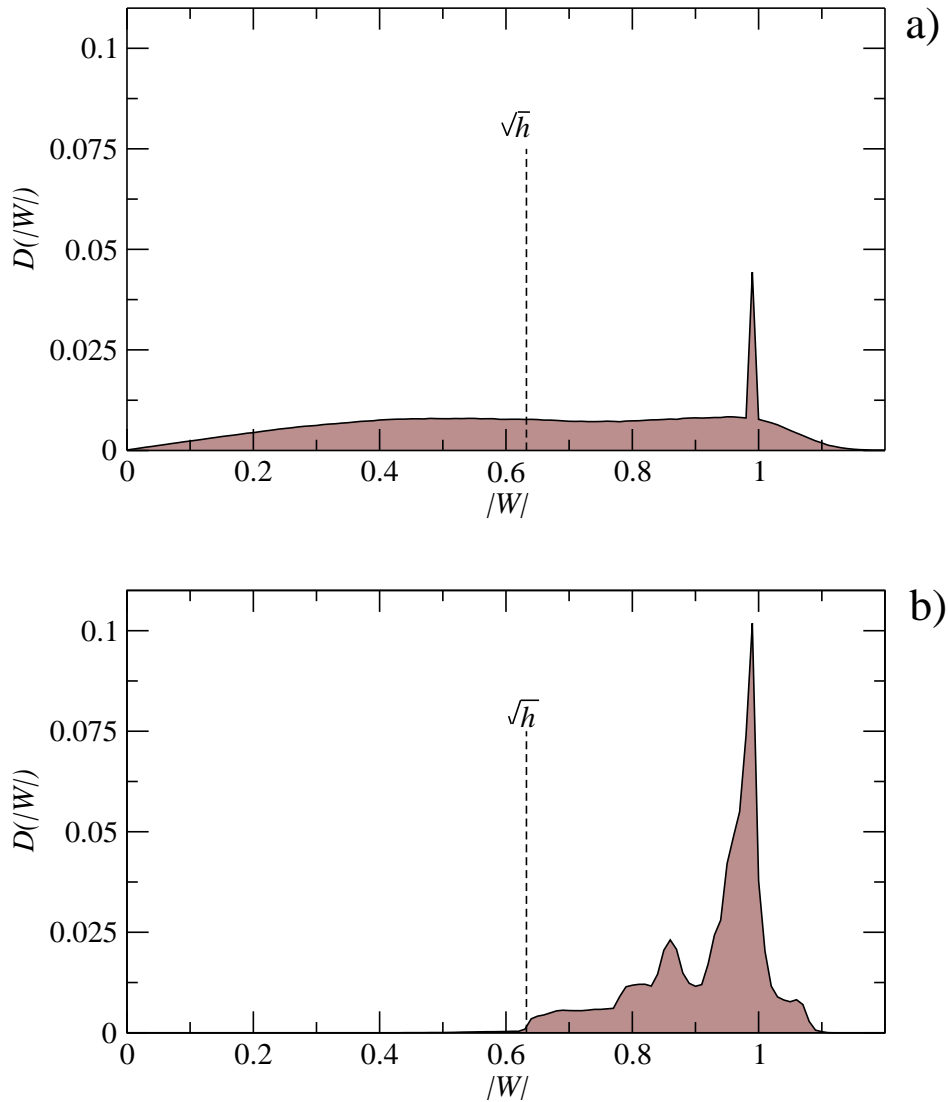}
			\caption{Distributions of the oscillation amplitude $|W|$  before (a) and after (b) the learning. The aim of the evolution is to minimize the number of oscillators with the amplitudes smaller than $\sqrt{h} \approx 0.63$.} 
			\label{fig_KT_3d} 
		\end{center}
	\end{figure}

The effect of learning is further illustrated in Fig. \ref{fig_KT_3d} where the distribution of the oscillator amplitudes $D(|W|)$ is plotted for the initial and the final states. Initially, many oscillators have small oscillation amplitudes below $\sqrt{h}$. After the learning, the oscillators move away from the center of the plane, avoiding the amplitude death. 

\section{Discussion}
\label{sec_conclusions} 

Dynamical systems can adapt their internal organization, e.g. the connection weights in a network, in such a way that a desired target dynamics is reproduced. The adaptation is performed through a built-in feedback loop.  In this loop,  the actual and the target performances of a system are compared and the resulting error signal is used to modify the structure of the system. Thus, through an autonomous evolution process, the system learns to generate a definite output.

In this study, an improved version of the previously proposed scheme with delayed feedbacks \cite{kaluza_pre2014} was considered and a novel autonomous learning scheme with inertial feedbacks was formulated. Numerical tests for selected optimization problems have shown that both methods produce satisfactory results. Since the proposed methods are empirical, it would be important to check them for a broader range of applications too.

Although only their computer simulations have been so far performed, the proposed methods may also be implemented, partially or completely,  at the "hardware" level.

The guided dynamical system (1) can be of physical, chemical or biological origin. Its equation of motion and even a complete set of its dynamical variables may be unknown. Essentially, such a real-world system can represent a "black box" whose output signals are controlled by varying a parameter set.

The steering system (2) should monitor the outputs and control the parameters of the guided system. It is not necessary that this is done by a computer in a digital way. Instead, this system can represent, for instance, an electronic circuit where the dynamics corresponding to the equations (2) is physically reproduced. Note that, in old analog computers, even the systems with much more complex dynamics could be implemented by electronic means.

Moreover, the two components, steering and guided, can also be incorporated into one physical device.  Through an internal adaptation process, such a device would be able to generate the required outputs and to maintain these outputs despite large variations of environmental conditions or the occurrence of faults.

Finally, a question can be asked whether similar mechanisms are already used by biological systems, even so small as a single living cell. In the latter case, the steering slow component can be, for example, the genetic subsystem, whereas the guided fast component can represent the metabolic subsystem of a cell. This is an interesting topic for further research.  

One of the authors (P.K.) acknowledges financial support from SeCTyP-UNCuyo (project M009 2016-2018) and from CONICET (PIP 11220150100013),
Argentina.

\end{document}